# Observation of two-species vortex lattices in a mixture of mass-imbalance Bose and Fermi superfluids

Xing-Can Yao[1,2,3,4], Hao-Ze Chen[1,2,3], Yu-Ping Wu[1,2,3], Xiang-Pei Liu[1,2,3], Xiao-Qiong Wang[1,2,3], Xiao Jiang[1,2,3], Youjin Deng[1,2,3], Yu-Ao Chen[1,2,3], & Jian-Wei Pan[1,2,3,4]

[1]Shanghai Branch, National Laboratory for Physical Sciences at Microscale and Department of Modern Physics, University of Science and Technology of China, Hefei, Anhui 230026, China

[2]CAS Center for Excellence and Synergetic Innovation Center in Quantum Information and Quantum Physics, University of Science and Technology of China, Shanghai, 201315, China

[3]CAS-Alibaba Quantum Computing Laboratory, Shanghai, 201315, China

[4]Physikalisches Institut, Ruprecht-Karls-Universität Heidelberg, Im Neuenheimer Feld 226，69120 Heidelberg, Germany

The superfluid mixture of interacting Bose and Fermi species is a remarkable many-body quantum system. Dilute degenerate atomic gases, especially for two species of distinct masses, are excellent candidates for exploring fundamental features of superfluid mixture. However, producing a mass-imbalance Bose-Fermi superfluid mixture, providing an unambiguous visual proof of two-species superfluidity and probing inter-species interaction effects remain challenging. Here, we report the realization of a two-species superfluid of lithium-6 and potassium-41. By rotating the dilute gases, we observe the simultaneous existence of vortex lattices in both species, and thus present a definitive visual evidence for the simultaneous superfluidity of the two species. Pronounced effects of the inter-species interaction are demonstrated through a series of precision measurements on the formation and decay of two-species vortices. Our system provides a new platform for studying novel macroscopic quantum phenomena in vortex matter of interacting species.

# I. Introduction

Superfluidity is one of the most intriguing macroscopic quantum phenomena manifesting itself in various fascinating effects, such as dissipationless flow via obstacles [1, 2], quantized vortices [3], and metastable persistent currents [4]. Quantized vortices, topological defects of quantized angular momentum [5], are a direct consequence of macroscopic superfluid wavefunction. In a superfluid, these vortices crystallize an Abrikosov lattice pattern [6] that minimizes vortex-vortex interaction. After the celebrated experimental realization of Bose-Einstein condensate [7, 8] (BEC), a series of studies were immediately devoted to creating vortices in BECs [9-11]. Several theoretical works emphasize that vortex lattices, directly visualizable in experiment, can serve as a conclusive evidence of superfluidity [12, 13]. For a dilute fermionic $^6$Li gas, such a definite visual evidence of superfluidity has been presented by the observation of vortex lattices [14]. Tremendous theoretical and experimental efforts have been devoted to understanding the behaviors of quantized vortices in BEC and quantum degenerate Fermi gases [14-19]. These studies have yielded a wealth of information and insights into diverse superfluid phenomena.

A superfluid mixture, where both components are superfluid, is a longstanding research topic in quantum physics. A mixture of bosonic $^4$He and fermionic $^3$He superfluid is predicted to exhibit either s-wave or topological p-wave Cooper pairs in the $^3$He component [20]. However, superfluid transition in the $^3$He component has not been detected even when the helium mixtures have been cooled down to about 100 μK [21]. In cold atom physics, it is now possible to experimentally produce mixtures of degenerate bosonic and fermionic dilute gases [22-24]. Very recently, a significant progress [25] was made in a mixture of two lithium isotopes—$^6$Li and $^7$Li, where the evidence of double superfluidity was provided by the occurrence of $^7$Li BEC and the density-profile difference of the two imbalanced $^6$Li spins. Further experimental studies on collective oscillations [25] and counterflowing dynamics [26] are consistent with theoretical predictions [27, 28]. However, the demonstration of two-species vortices, which not only provides a conclusive evidence of double superfluidity, but also gives more insights into fundamental properties of double superfluid, is still unavailable. It is thus an important and demanding task to observe the simultaneous existence of vortex lattices in the two superfluids, and further to study their rich behaviors arising from vortex-vortex interactions.

Moreover, in contrast to a double superfluid of the two isotopes, a double superfluid of two distinct species, if experimentally realized, may exhibit more pronounced interaction effects due to large mass-imbalance [29, 30]. These Bose-Fermi interactions, without which a double superfluid is just of two independent superfluids, play an essential role in the emergence of many fascinating quantum phenomena [31], including Bose-Fermi dark-bright solitons [32], and polaronic atom-trimer continuity [33] etc. The two-species superfluid of dilute atomic gases can also provide a powerful tool for simulating real materials and beyond—e.g., dense quantum chromodynamics matter [34]. However, due to the mass imbalance, producing a two-species Bose-Fermi superfluid remains challenging.

Here we report on the generation of a two-species superfluid of fermionic $^6$Li and bosonic $^{41}$K atoms in an optical dipole trap, which have a mass ratio of about seven. Moreover, we overcome several technical challenges and produce a double-vortex matter in which both superfluid components simultaneously exhibit vortex lattices. Finally, a series of precise measurements on the formation and decay of the two-species vortices reveal pronounced effects of the inter-species interaction. We also stress that the vortex-vortex and the inter-species interactions may induce many unconventional behaviors in such a vortex matter of interacting species, remaining to be explored both theoretically and experimentally.

Producing a two-species superfluid of $^6$Li and $^{41}$K atoms ($2S_{Li-K}$) and simultaneously creating vortex lattices in both species represent various significant technical challenges. Firstly, due to the unresolved D2 excited hyperfine structure of both $^6$Li and $^{41}$K atoms, the conventional sub-Doppler cooling techniques become ineffective [35, 36]. We adopt an advanced laser-cooling scheme, where a sub-Doppler cooling in gray molasses is implemented for $^{41}$K while a near-detuned Doppler cooling is applied for $^6$Li. In addition, a two-stage evaporative cooling strategy is employed where $^6$Li is first sympathetically cooled by $^{41}$K in the optically-plugged magnetic trap and then $^{41}$K is sympathetically cooled by $^6$Li in the optical dipole trap. Such an evaporation cooling procedure is essential to generate a 2S$_{Li-K}$ with both large atom numbers and tunable atomic ratio. Secondly, the large mass-imbalance can induce inhomogeneous spatial overlap or even separation of the two superfluids due to the gravitational sag [37]. We utilize a disk-like optical dipole trap to confine the cloud, where two elliptical laser beams with an aspect ratio of 4:1 are crossed perpendicularly (See Fig. 1b). The spatial overlap of the two superfluids is obtained due to the high trapping frequency in the gravity direction. Meanwhile, the disk-like trap provides a considerable large trap volume and relatively low geometric mean trapping frequency, enabling a large production and long lifetime of 2S$_{Li-K}$. More importantly, the trap exhibits a homogeneous trapping potential and a rotational symmetry along the *z* direction, crucial for the simultaneous nucleation of vortices. In comparison with the standard single- or crossed-beam dipole traps, the disk-like trap is unique that it simultaneously enjoys the aforementioned advantages. In short, owing to the exquisite control of evaporation process and the unique design of the disk-like trap, we not only produce a large two-species superfluid as well as two-species vortex lattices, but also have a wide range of tunable parameters.

## II. Two-species Bose-Fermi superfluid

Our first goal is to produce a two-species Bose-Fermi superfluid. After the laser cooling phase, the cold gas mixture is magnetically transported to a dodecagonal glass-cell of good optical access and ultra-high vacuum environment (see Fig. 1a). The first-stage evaporative cooling is applied to $^{41}$K in an optically-plugged magnetic trap by driving $|F=2,m_F=2\rangle$ to $|F=1,m_F=1\rangle$ transition of $^{41}$K, while $^6$Li is cooled sympathetically. By optimizing the radio-frequency (RF), we achieve a near-degenerate mixture of $5\times10^{7}$ $^6$Li and $1\times10^{7}$ $^{41}$K atoms. Then, the mixture is transferred into a cigar-shaped dipole trap (wavelength 1064 nm, 1/e$^2$ radius 35 μm and maximum laser power 10 W), followed by two Landau-Zener sweeps which immediately prepare both species at their lowest hyperfine states, labeled $|1_{Li}\rangle$ and $|1_K\rangle$. Next, a half-to-half spin mixture of the two lowest hyperfine states of $^6$Li is prepared by using successive RF sweeps at a magnetic field of 870 G [38]. At this field, the background scattering length of $^{41}$K is 60.5$a_0$ ($a_0$ is the Bohr radius) and that of $^6$Li-$^{41}$K is 60.7$a_0$ [39, 40]. The spin mixture of $^6$Li is located at the Bardeen-Cooper-Schrieffer (BCS) side of Feshbach resonance[41] (scattering length -12580$a_0$). Due to the unitary limited collisional rate between fermionic spins, the forced evaporative cooling of $^6$Li is rather efficient and $^{41}$K is sympathetically cooled.

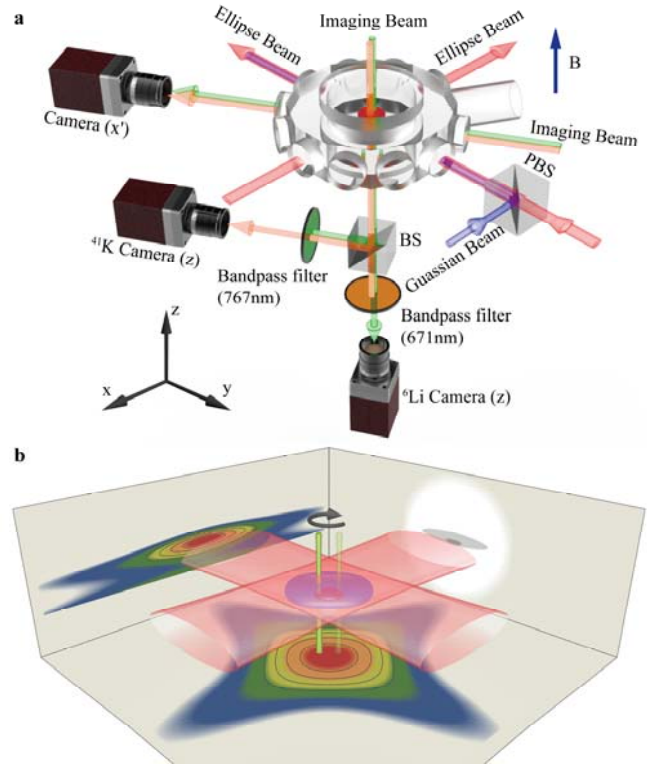

**Figure 1 | Sketch of experimental set-up (a) and specially designed disk-like trap (b). a,** An elliptical beam (light red) is superimposed with a Gaussian beam (blue) by a polarizing beam splitter, and then crossed with another elliptical beam (light red) at 90°. The dichromatic imaging beams are used for simultaneous absorption imaging of $^6$Li (green) and $^{41}$K (orange) atoms along the $z$ axis. The imaging beams after a beam splitter are respectively filtered by 671 nm and 767 nm bandpass filters, and then detected by the $^6$Li and $^{41}$K EMCCD cameras. An additional EMCCD camera is used for imaging the two species along $x'$ axis. **b,** Two elliptical Gaussian beams (light red) with an aspect ratio of 4:1 are crossed perpendicularly, forming a disk-like optical potential. The $^6$Li and $^{41}$K clouds are denoted by purple and red ellipsoid. They are rotated by two laser beams (green) with a wavelength of 532 nm for the creation of vortices. The contour plots represent cross-sections of the trapping potential in the $xz$ and $xy$ planes, respectively. The grayscale image shows a simulated density distribution of the two-species superfluid in the $yz$ plane at the experimental trap geometry.

After 0.5s of evaporation, a disk-like trap (wavelength 1064 nm, maximum laser power of each beam 1.1 W, $1/e^2$ axial (radial) radius 48 μm (200 μm)) is turned on slowly while the laser power of the cigar-shaped trap is ramped down in 0.3 s. Subsequent evaporation is achieved by exponentially lowering the laser intensity to 120 mW in 2 s. At the end of evaporation, the trap is held for 1 s to ensure fully thermal equilibrium between both species. The axial trap frequency of $^6$Li ($^{41}$K) is 237 Hz (85 Hz), which is strong enough to attain a complete overlap of two superfluids (see Fig. 1b). The horizontal confinement is provided by the combination of magnetic curvature and optical dipole trap, resulting in a radial trap frequency of 40 Hz (20 Hz) for $^6$Li ($^{41}$K). The spatial overlap is checked by imaging along the $x'$ axis (Fig. 1a). Further, a specially designed imaging setup along the $z$ axis is employed to simultaneously probe the two species (Fig. 1a). An imaging resolution of 2.2 μm (2.5 μm) at 671 nm (767 nm) is gained by a high numerical aperture objective, which is very important for observing vortices.

A two-species Bose-Fermi superfluid is finally achieved at 870 G with more than $1.8 \times 10^5$ $^{41}$K atoms in a BEC (condensate fraction ≥90%) and $1.5 \times 10^6$ $^6$Li atoms at 7% Fermi temperature [25]. The Thomas-Fermi radii of the $^{41}$K component are about 5.5 μm axially and 22.3 μm radially, and the $^6$Li component has a Fermi radii of about 21 μm axially and 121 μm radially. The superfluid mixture is very stable, with a 1/e lifetime 12.0±0.7 s for $^{41}$K and 19±1 s for $^6$Li. Figure 2 shows a series of in-situ images, in which the magnetic field is varied and thus the ultracold spin mixture of $^6$Li experiences a crossover from a molecular BEC [42] to a BCS superfluid [43]. It can be seen that the radius of the $^6$Li cloud increases gradually from the BEC to the BCS side, while the $^{41}$K cloud size remains unchanged. The geometric centers of the two species are perfectly overlapped in the $xy$ plane, and a full overlap in the gravity direction is also achieved (see Fig. 2). The disk-like trap exhibits an approximate rotational symmetry along the $z$ axis in the center, as shown by the nearly circular shape of the $^{41}$K component. The squared shape formed by the edges of the $^6$Li superfluid reflects an imperfection of the rotational symmetry in a larger scale, but might give rise to intriguing vortex structures.

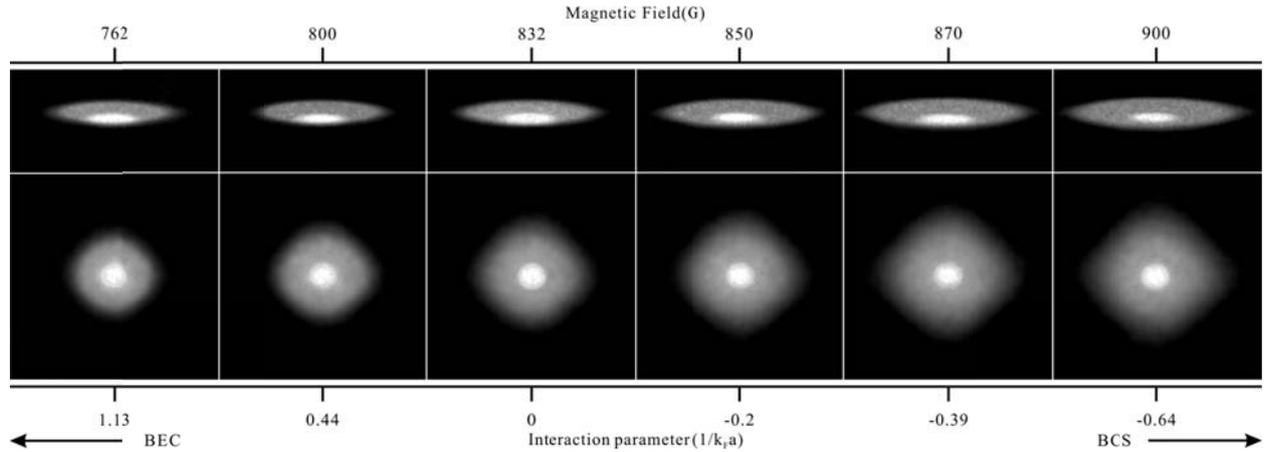

**Figure 2 | In-situ images of two-species Bose-Fermi superfluid of $^{41}$K and $^{6}$Li.** The top- and bottom-row images respectively show the integrated optical density distributions of the two-species superfluid along the $x'$ and $z$ direction, as the magnetic field is varied from the BEC to the BCS side of the $^{6}$Li. The $^{6}$Li component has an inverse Fermi momentum of $1/k_F \approx 0.26$ μm. To take an in-situ image, we prepare the superfluid mixture at 870 G, ramp the magnetic field to the desired value (from 762 G to 900 G) in 100 ms, and hold the trap for another 100 ms before imaging. Each picture is obtained by carefully superimposing two images that are taken separately in two experimental cycles (top) or simultaneously in a single cycle (bottom). It is seen that the $^{41}$K superfluid (bright region) is fully buried in the $^{6}$Li superfluid cloud. The viewing scope of top (bottom) images is 416×208 μm$^2$ (416×416 μm$^2$).

### III. Observation of vortex lattices

Rotational symmetry is an important ingredient for generating vortices, especially for fermionic superfluid. In experiment, we utilize two custom-designed anamorphic prism pairs to create the elliptical beams with minimum optical aberrations. Further, special effort is devoted to aligning the trap center with respect to the saddle point of magnetic curvature. The cloud is rotated about its symmetry axis (see Fig. 1b) by a blue-detuned laser beam (wavelength 532 nm, $1/e^2$ radius 19 μm). A two-axis acousto-optic deflector is used to regulate the motion of the stirring beam, where a two-beam pattern with tunable separation is rotated symmetrically with a variable angular frequency [10]. Nevertheless, the imperfection of large-scale rotational symmetry in our system might inhibit formation of $^{6}$Li vortices. Another challenge comes from probing vortices of either species, for which slow axial and rapid radial expansions are preferred to avoid vortex-core bending and image blurring. Unfortunately, the expansion of atoms confined in the disk-like trap experiences an exactly opposite condition.

We first produce vortex lattices for the single-species superfluids. For the fermionic $^6$Li superfluid at 870 G, the rotating beams (laser power 2 mW, beam separation 76 μm) are ramped up to stir the cloud with a frequency of 26 Hz during the last second of evaporation. They are then adiabatically turned off, and the trap is held for 2 s. To probe vortices [14], the disk-like trap is suddenly switched off and the cloud is expanded in the residual magnetic curvature. During the first 3 ms of expansion the magnetic field is rapidly ramped to 740 G, and the cloud is further expanded for another 17 ms at this field. A molecular absorption imaging is then taken, where the probe laser resonantly excites atoms in state $\left|1_{Li}\right\rangle$ [44]. A clear Abrikosov lattice pattern[6] of 19 $^6$Li vortices is obtained (Fig. 3a), with a nearly 60% visibility of the vortex core. By using the same procedure, a regular lattice of 21 vortices is also observed (Fig. 3b) in the molecular BEC of $^6$Li at 762 G. For the bosonic $^{41}$K superfluid at 870 G, we gain a highly ordered triangular lattice pattern of 20 vortices with a visibility about 70% (Fig. 3c).

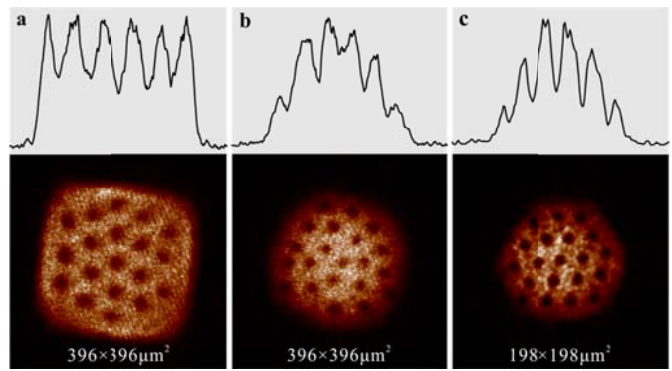

**Figure 3 | Vortex lattices in single-species superfluids. a,** $^6$Li vortices at 870 G (BCS), **b,** $^6$Li vortices at 762 G (BEC), and **c,** $^{41}$K vortices at 870 G. The curves show the integrated density profiles of 11μm-wide-cut through the two-dimensional images below. In all the three cases, the triangular-lattice geometry is clearly observed.

In the two-species superfluid, nucleation of vortex lattices is more complex, arising from several factors like the inter-species interaction and the variation of population ratio. Furthermore, the optimal stirring parameters for the bosonic and fermionic components are rather distinct due to the difference in size, density and trap frequency. We explore a wide range of combinations of stirring parameters and successfully find feasible ones. For creation of two-species vortices, the superfluid mixture is prepared at magnetic field 870 G. The cloud is rotated during the last 600 ms of evaporation, and then equilibrated for another 2 s. Absorption imaging for $^6$Li and $^{41}$K atoms are then simultaneously performed.

A series of highly visible two-species vortex lattices are observed (Fig. 4), as stirring parameters (beam separation, rotating frequency and stirring-laser power) are varied. Figures 4a and 4d show a two-species vortex lattice, consisting of an ordered triangular lattice of 20 $^{41}$K vortices and 2 symmetric $^6$Li vortices. A regular lattice of 6 $^6$Li vortices is also obtained (Fig. 4c), but only 4 vortices survive in a small $^{41}$K condensate (Fig. 4f), due to a significantly weakened effective trapping force for $^{41}$K. Figures 4b and 4e contain a diamond of 4 $^6$Li vortices and a triangular lattice of 14 $^{41}$K vortices, respectively.

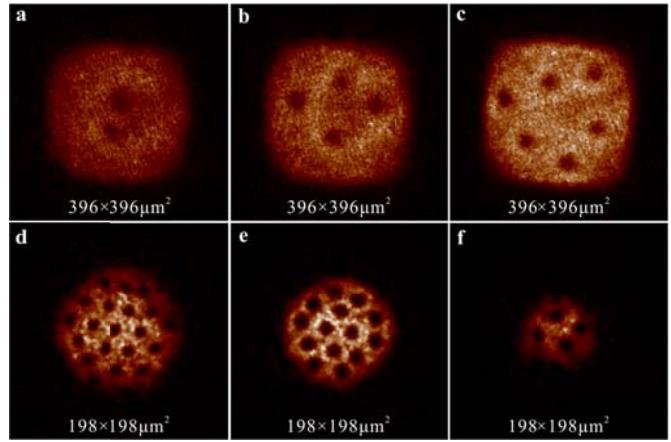

**Figure 4 | Vortex lattices in two-species superfluid.** The top (bottom) images are for the $^6$Li ($^{41}$K) component, and the two images in each column, taken simultaneously in an experimental cycle, are for the same set of stirring parameters. **(a, d)** is for (beam separation 28 μm, laser power 0.75 mW, rotating frequency 26 Hz), **(b, e)** for (52 μm, 0.75 mW, 18 Hz), and **(c, f)** for (76 μm, 1.5 mW, 22 Hz).

The observed vortex lattices confirm the advantages of the disk-like trap in the study of superfluid mixture of two dramatically distinct species. More importantly, the simultaneous formation of vortex lattices in both components provides a definitive visual evidence for the peculiar state of Bose-Fermi double superfluidity. It is also noted that depending on the strength of inter-species interactions, the general structure of two-species vortex lattices may undergo interesting phase transitions.

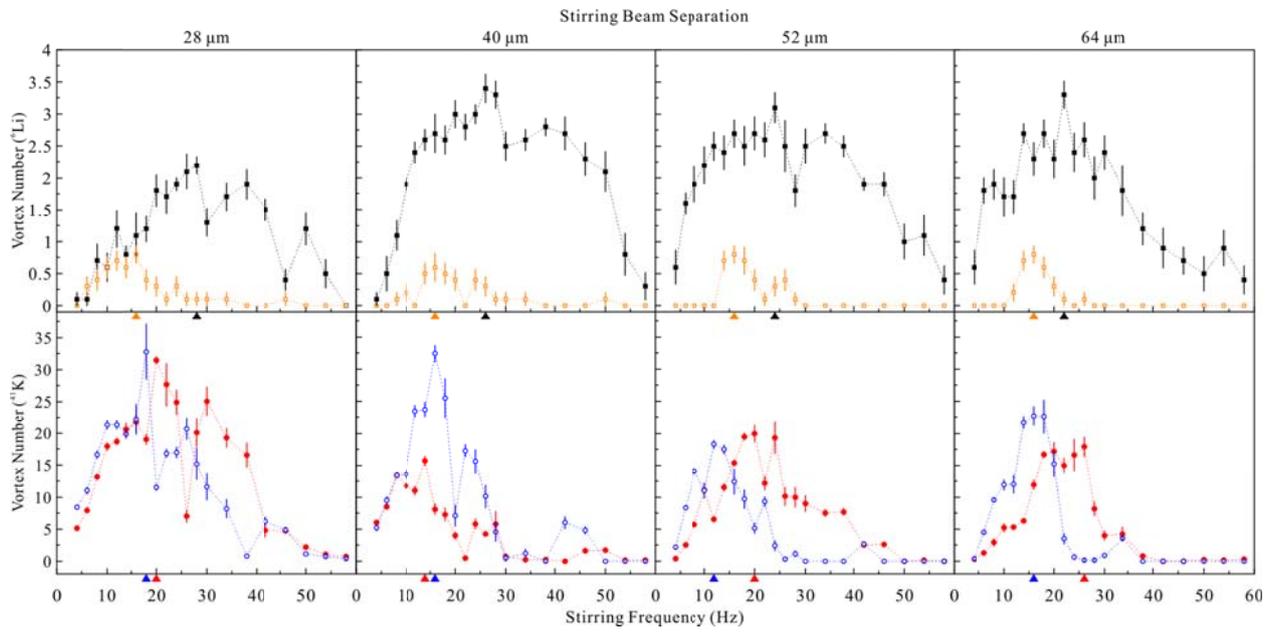

**Figure 5 | Comparison of vortex numbers in single- and two-species superfluids.** For a given set of stirring parameters, three comparison experiments are respectively performed for the superfluid mixture, and the single-species superfluid of $^6$Li or $^{41}$K. Vortex number is counted and statistically averaged over 10 measurements with standard error margin being calculated. The numbers of $^6$Li vortices in the two- (single-) species superfluids are denoted by the filled black (open orange) squares and plotted on the top row; the $^{41}$K vortex numbers are analogously shown as the filled red (open blue) circles on the bottom row. For each column of plots, the beam separation is fixed, while the rotation frequency is varied. Optimal frequencies for creating a maximal number of vortices are indicated by the triangles on the horizontal axis, filled with the same color of the data points. The inter-species interaction effects can be clearly seen from the dramatic difference of the $^6$Li vortex numbers (top row), over a wide range of rotation frequency.

## IV. Formation and decay of two-species vortices

In a superfluid mixture of the two lithium isotopes, it is observed [25] that although too weak to alter the density profiles, the Bose-Fermi interaction affects the collective dynamics of the

system—e.g., inducing a beating behavior in the dipole-oscillation of the bosonic component. In our system, the masses of the $^{41}$K and $^{6}$Li atoms are sufficiently imbalanced, and thus the Bose-Fermi and Bose-Bose couplings are significantly different. As theoretically predicted [30,45], we observe that the repelling from the $^{41}$K superfluid component leads to a density-profile depression in the center of the $^{6}$Li component. Such an alternation is also clearly seen in the presence of vortices (Figs. 4a and 4b). This implies that the inter-species interaction might play an important role in our system—e.g., on the formation and decay of vortex lattices.

We perform a series of comparison experiments on the formation of vortices, in which the same set of stirring parameters (stirring time 0.6 s, equilibrium time 2 s, laser power 0.75 mW) is respectively applied to the two-species superfluid as well as the single-species superfluid of $^{6}$Li or $^{41}$K. Furthermore, by adjusting the loading parameters, we also make sure that the temperature and the atom numbers of $^{6}$Li and $^{41}$K are approximately identical in the single- and the two-species superfluids. Precision measurements are carried out for the average numbers of vortices as a function of rotating frequency. Figure 5 shows four sets of experimental results for different beam separations.

Pronounced effects of the Bose-Fermi coupling are revealed, particularly from the comparison of $^{6}$Li vortex numbers (the top row of Fig. 5). Strikingly, in the two-species superfluid the $^{6}$Li vortex number is greatly increased and the range of vortex-generation rotating frequency is largely widened. For some ranges of stirring parameters (e.g., 64 um of beam separation and 46 Hz of rotating frequency), no vortex can be formed in the single-species superfluid of $^{6}$Li or $^{41}$K, but in the two-species superfluid, the $^{6}$Li vortices are unexpectedly created while the $^{41}$K vortex is still absent. In addition, we observe that as beam separation is increased, the optimal rotating frequency corresponding to a maximal number of $^{6}$Li vortices has a spectacular downshift (from 28 Hz to 22 Hz) in the two-species superfluid while it remains unchanged in the single-species superfluid. These behaviors strongly show that the $^{41}$K component acts as an important role in the formation and behavior of $^{6}$Li vortices in the two-species superfluid. The coupling effect on generating $^{41}$K vortices is less pronounced, but still visible in the shift of the optimal rotating frequency. We stress that these interaction-induced effects can be quantitatively characterized, as depicted in Fig. 5.

Theoretical insights are urgently desired to reveal the underlying mechanisms, possibly beyond the scope of mean-field descriptions. It seems plausible that the interaction would lead to both energy and/or angular-momentum transfer between the bosonic and fermionic components. However, it remains an open and important research topic whether and how the transfers happen. Quantitative models are particularly needed to account for our experimental results in Fig. 5. With deeper understandings, one might expect to predict a variety of novel macroscopic quantum phenomena.

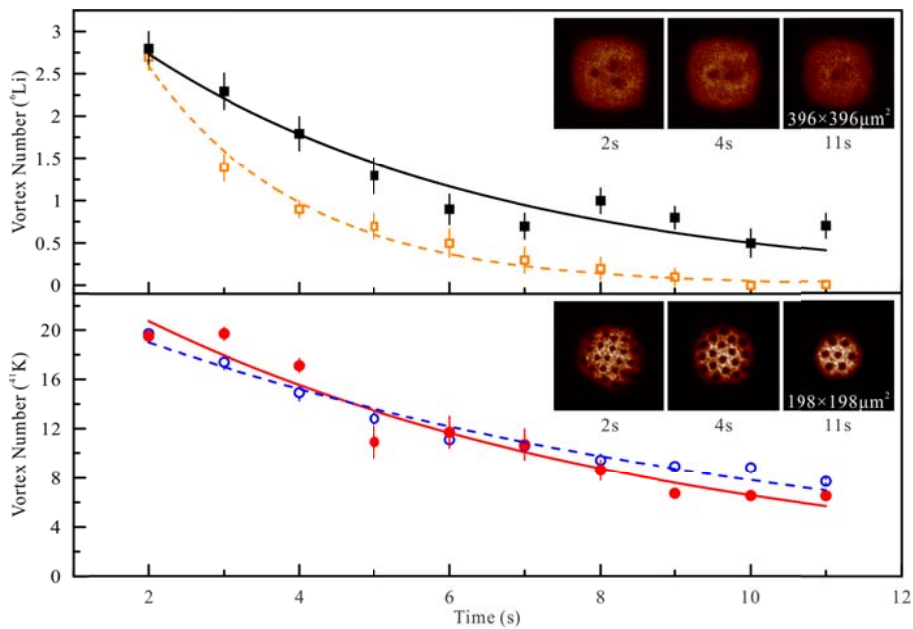

**Figure 6 | Decay of vortices in single- and two- species superfluids.** As in Fig. 5, three comparison experiments are carried out with the initial $^6$Li ($^{41}$K) vortex numbers being set approximately identical (see text). Also as in Fig. 5, symbols are used in the same way—e.g., the filled black squares depict the decay of $^6$Li vortices in the superfluid mixture, and each data point has a standard error bar and is gained by averaging over 10 measurements. The curves are obtained by fitting the experimental data to an exponential decay. The insets show typical images of vortices in the two-species superfluid, suggesting the survival of both $^6$Li and $^{41}$K vortices after 11 s.

We also carry out comparison experiments on the decay of vortices. By adjusting the stirring parameters, the average numbers of $^6$Li ($^{41}$K) vortices are initialized to be nearly the same in the single- and two-species superfluids. Figure 6 shows the results of the initial vortex numbers (2.8±0.2, 19.5±0.5) for $^6$Li and $^{41}$K. Impressively, the $^6$Li vortices seem more stable in the two-species superfluid than in the single-species superfluid, with a 1/e lifetime $\tau_2$=4.7±0.5 s twice longer than $\tau_1$=2.1±0.2 s. In contrast, the $^{41}$K vortex lifetimes are $\tau_2$=6.9±0.5 s for the superfluid mixture and $\tau_1$=9.1±0.6 s for the single-species superfluid. Again, we argue that this behavior might be attributed to the possible angular-momentum transfer between the two species in the superfluid mixture.

## V. Conclusion

We have employed a combination of state-of-the-art techniques, including advanced laser-cooling scheme and specially designed disk-like optical trap etc., and produced a two-species Bose-Fermi superfluid. Our work represents an important progress in experimentally realizing a superfluid mixture of two distinct species, a long-standing goal in cold-atom physics. According to our knowledge, the obtained superfluid mixture is also the largest one, containing more than $1.8 \times 10^5$ $^{41}$K atoms and $1.5 \times 10^6$ $^6$Li atoms. Vortex lattices—quantized topological excitations of superfluid—are successfully generated and directly visualized in both the single- and the two-species superfluids. The large mass-imbalance of $^{41}$K and $^6$Li may lead to pronounced effects of inter-species interaction, which are clearly revealed in the formation and decay of two-species vortices and request deep theoretical understandings. With a wide range of tunable parameters like population ratio of two species, magnetic field controlling the BEC-BCS crossover and stirring parameters, many static and dynamic properties of such Bose-Fermi superfluid mixtures can be explored, particularly those due to inter-species coupling, vortex-vortex interaction, and their interplay.

# Reference

1. C. Ramanm, *et al.*, Phys. Rev. Lett., **83,** 2502 (1999).

2. R. Onofrio, *et al.*, Phys. Rev. Lett., **85,** 2228 (2000).

3. M. M. Salomaa, G. E. Volovik, Rev. Mod. Phys., **59,** 533 (1987).

4. S. Beattie, S. Moulder, R. J. Fletcher, Z. Hadzibabic, Phys. Rev. Lett., **110,** 025301 (2013).

5. E. M. Lifshitz, L. P. Pitaevskii, Statistical Physics, Part 2 (Pergamon, Oxford, 1980).

6. A. A. Abrikosov, Sov. Phys. JETP, **5,** 1174 (1957).

7. M. H. Anderson, *et al.*, Science, **269,** 198 (1995).

8. K. B. Davis, *et al.*, Phys. Rev. Lett., **75,** 3969 (1995).

9. M. R. Matthews, *et al.*, Phys. Rev. Lett., **83,** 2498 (1999).

10. K. W. Madison, F. Chevy, W. Wohlleben, J. Dalibard, Phys. Rev. Lett., **84,** 806 (2000).

11. J. R. Abo-Shaeer, C. Raman, J. M. Vogels, W. Ketterle, Science, **292,** 476 (2001).

12. M. Rodriguez, G. S. Paraoanu, P. Törmä, Phys. Rev. Lett., **87,** 100402 (2001).

13. L. Pitaevskii, S. Stringari, Science, **298,** 2144 (2002).

14. M. W. Zwierlein, J. R. Abo-Shaeer, A. Schirotzek, C. H. Schunck, W. Ketterle, Nature, **435,** 1047 (2005).

15. I. Coddington, P. Engels, V. Schweikhard, E. A. Cornell, Phys. Rev. Lett., **91,** 100402 (2003).

16. D. V. Freilich, D. M. Bianchi, A. M. Kaufman, T. K. Langin, D. S. Hall, Science, **329,** 1182 (2010).

17. A. L. Fetter, Rev. Mod. Phys., **81,** 647 (2009).

18. M. W. Zwierlein, A. Schirotzek, C. H. Schunck, W. Ketterle, Science, **311,** 492 (2006).

19. S. Giorgini, L. P. Pitaevskii, S. Stringari, Rev. Mod. Phys., **80,** 1215 (2008).

20. J. Rysti, J. Tuoriniemi, A. Salmela, Phys. Rev. B, **85**, 134529 (2012).

21. J. Tuoriniemi, *et al.*, Journal of Low Temperature Physics, **129,** 531 (2002).

22. Z. Hadzibabic, *et al.*, Phys. Rev. Lett., **88,** 160401 (2002).

23. C. Silber, *et al.*, Phys. Rev. Lett., **95,** 170408 (2005).

24. J. Goldwin, *et al.*, Phys. Rev. A, **70,** 021601 (2004).

25. I. Ferrier-Barbut, *et al.*, Science, **345,** 1035 (2014).

26. M. Delehaye, *et al.*, Phys. Rev. Lett., **115,** 265303 (2015).

27. W. Zheng, H. Zhai, Phys. Rev. Lett., **113,** 265304 (2014).

28. M. Abad, A. Recati, S. Stringari, F. Chevy, The European physical Journal D, **69,** 126 (2015).

29. R. Zhang, W. Zhang, H. Zhai, P. Zhang, Phys. Rev. A, **90,** 063614 (2014).

30. T. Ozawa, A. Recati, M. Delehaye, F. Chevy, S. Stringari, Phys. Rev. A, **90,** 043608 (2014).

31. A. B. Kuklov, B. V. Svistunov, Phys. Rev. Lett., **90,** 100401 (2003).

32. M. Tylutki, A. Recati, F. Dalfovo, S. Stringari, New Journal of Physics, **18,** 053014 (2016).

33. Y. Nishida, Phys. Rev. Lett., **114,** 115302 (2015).

34. K. Maeda, G. Baym, T. Hatsuda, Phys. Rev. Lett., **103,** 085301 (2009).

35. P. M. Duarte, *et al.*, Phys. Rev. A, **84,** 061406 (2011).

36. G. Modugno, *et al.*, Science, **294,** 1320 (2001).

37. R. Pires, *et al.*, Phys. Rev. Lett., **112,** 250404 (2014).

38. K. E. Strecker, G. B. Partridge, R. G. Hulet, Phys. Rev. Lett., **91,** 080406 (2003).

39. S. Falke, *et al.*, Phys. Rev. A, **78,** 012503 (2008).

40. E. Tiemann, *et al.*, Phys. Rev. A, **79,** 042716 (2009).

41. G. Zürn, *et al.*, Phys. Rev. Lett., **110,** 135301 (2013).

42. S. Jochim, *et al.*, Science **302,** 2101 (2003).

43. M. Bartenstein, *et al.*, Phys. Rev. Lett., **92,** 120401 (2004).

44. G. B. Partridge, W. Li, R. I. Kamar, Y. Liao, R. G. Hulet, Science, **311,** 503 (2006).

45. K. Mølmer, Phys. Rev. Lett., **80,** 1804 (1998).

**Acknowledgements**

We appreciate helpful discussions with H. Zhai. This work has been supported by the NSFC of China, the CAS, and the National Fundamental Research Program (under Grant No. 2013CB922001). X.-C. Y. acknowledges support from the Alexander von Humboldt Foundation.